\documentclass[english]{article}
\usepackage{graphicx,color}
\usepackage{amsmath,amssymb}
\usepackage{authblk}
\def\bc{\begin{center}}
\def\ec{\end{center}}

\def\beq{\begin{equation}}
\def\eeq{\end{equation}}

\def\d{\downarrow}
\def\u{\uparrow}

\begin{document}


\title{Monitoring of quantum walks with weak measurements}

\author[1,2]{Klaus Ziegler}
\author[3]{Tim Heine}
\author[4]{Sabine Tornow}
\affil[1]{Institut für Physik, Universität Augsburg, D-86135 Augsburg, Germany}
\affil[2]{Physics Department, New York City College of Technology,\\
The City University of New York,
Brooklyn, NY 11201, USA\\}
\affil[3]{Institute of Quantum Technologies, German Aerospace Center (DLR), Wilhelm-Runge-Str. 10, 
D-89081 Ulm, Germany}
\affil[4]{Research Institute CODE, University of the Bundeswehr Munich, D-81739 Munich, Germany
}

\maketitle

Measurements 
can be used to monitor the evolution of quantum systems and can give rise to quantized return 
statistics. It is known that the mean return time is quantized for strong monitoring through the
winding number of the monitored quantum state.
We discuss that under coherent weak monitoring, implemented via ancilla coupling, the mean return 
time of a quantum walk obeys a scaling relation with respect to the measurement strength.
An analog scaling relation was previously found for random-time monitoring, indicating that weak
and random-time monitoring have similar effects. We discuss how weak monitoring via ancilla coupling
is linked to the unitary evolution, and how this connection can be controlled by a convergent
perturbation theory. 

\section{Introduction}

To understand how measuring affects the return-time statistics is relevant both for fundamental questions 
about quantum walks and for applications where repeated measurements are used as diagnostic 
or control tools \cite{nielsen10}.
For several classes of continuous-time quantum walks, it has been shown that the mean number of 
projective detection attempts can exhibit a quantized structure governed by a winding number of a 
characteristic generating function \cite{Varbanov2008,gruenbaum13,yin19,kessler21,ziegler21,Dhar2017,Engelhardt2020}. In these cases, 
the mean return or transition time is determined only by how many distinct energy levels are visible to the 
detector, while it is not sensitive to the detailed distribution of spectral weights or the actual
values of the energy levels. 
These observations fit into a broader analysis of first-detected passage and transition times that 
combines renewal relations with spectral information \cite{Dhar2017,Engelhardt2020}.
Most of this work assumes strong measurements, modelled by rank-one projectors acting 
directly on the system \cite{Varbanov2008,gruenbaum13,yin19,kessler21,ziegler21,Dhar2017,Engelhardt2020}. 
In many physical implementations, however, indirect or weak measurements provide a more realistic description.
Examples include circuit-based architectures in which the system interacts with an ancilla that is subsequently 
measured, and continuous-monitoring scenarios, where the coupling to a detector is weak and information is 
acquired gradually over time \cite{nielsen10,gebhart20,dhar23}. In such settings the measurement is naturally 
described by Kraus operators or generalized measurements, with a tunable strength that interpolates between 
almost unitary evolution and the projective limit \cite{nielsen10}. Weak measurements of this type have been
studied in the context of measurement-induced phases in random circuits \cite{PhysRevB.101.104301,weakMIPT},
weak-measurement-induced geometric phases \cite{gebhart20,gebhart20weak}, quantum resetting under 
continuous measurement \cite{dhar23}, and quantum feedback and control protocols that explicitly use weak 
measurement backaction as a resource \cite{LloydSlotine2000}. A weak-measurement formulation of first-hitting 
times for quantum walks, with an explicit circuit-level description and implementations on superconducting quantum 
processors, has also been developed recently \cite{heine25}.

Despite this progress, the impact of weak monitoring on return laws and its connection to topologically
protected properties has not yet been systematically analyzed. In particular, it is not obvious whether the 
quantization of the mean return time in terms of a winding number survives when projective measurements 
are replaced by weak ones.
In this work
we consider the return-time statistics of a quantum walk that is monitored by repeated weak measurements at a
detector state. After each unitary step $U_\tau = \mathrm{e}^{-\mathrm{i}H\tau}$ over a fixed time $\tau$
an indirect measurement of tunable strength $\eta$ is performed. The combination of $U_\tau$ with the
measurement operator creates a monitored-evolution operator. Within this framework we derive a formal relation 
between the generating functions of the projectively monitored and weakly monitored return amplitudes, 
and use it to analyze the first-detection statistics. 
We will show that the mean return time $\langle t\rangle$, which depends 
on the strength of measurement $\eta$ ($0 < \eta \leq 1$), interpolates as $\langle t\rangle=\tau n_w/\eta$
between the unitary limit $\eta=0$ and the projective limit $\eta=1$.
$n_w$ is the winding number that counts the number of detector-visible energy levels 
\cite{gruenbaum13,yin19,ziegler21,heine25}. Thus, the mean return time is robust against weakening 
the measurement and remains determined solely by the underlying topology of the projective
measurement.

The structure of this paper is as follows: In Sect. \ref{sect:monitoring} we introduce monitoring by
indirect measurements. For this purpose we begin with the discussion of the unitary evolution
in Sect. \ref{sect:unitary}, followed by the definition and study of monitoring by weak measurements in
Sect. \ref{sect:monitor}. The concepts developed in these two sections are applied to the mean 
return time in Sect. \ref{sect:mrt}. In Sect. \ref{sect:variance} we analyze the variance of the return time 
and calculate it for the instructive example of a two-level system. 
The results of our analysis of monitoring by indirect measurements are discussed and
summarized and its connection with random-time measurement
is briefly discussed  in Sect. \ref{sect:discussion}.

\section{Monitoring by indirect measurements}
\label{sect:monitoring}

Strong measurement refers to an idealized, projective measurement, with no 
further disturbance to the system. Regarding a quantum algorithmic framework, where the unitary operator acts
on a finite register of qubits, a measurement is often realized by additional ancilla qubits that can be
understood as a weak measurement. In practice, such measurements affect the quantum evolution less 
than strong, projective measurements. In other words, these weak measurements indicate 
that the interaction between the system and the measurement apparatus is sufficiently small, resulting in a
balanced information gain per measurement and a minimal disturbance to the system. This may manifest as 
increased uncertainty in the measurement outcomes, which can resemble additional randomness.

A simple generic protocol to monitor the quantum walk indirectly was proposed in Ref.~\cite{heine25}.
In this case the measurement admits a linear shearing of the projection operator, created by weak
coupling to the ancilla through controlled RY gates. 
Hence, we map this protocol to the framework of quantum recurrence~\cite{gruenbaum13} with 
the coupling strength $\eta$ as a tunable parameter~\cite{gebhart20,dhar23,heine25}:
\beq
\label{def_proj}
\begin{pmatrix}
P_\eta \\
Q_\eta
\end{pmatrix}=\begin{pmatrix}
\sqrt{\eta} & 0 \\
\sqrt{1-\eta} & 1 
\end{pmatrix}\begin{pmatrix}
P \\
Q \\
\end{pmatrix}
\eeq
with the projection operators $P$, $Q$ and $P+Q={\bf 1}$. This interpolates between the unitary evolution
for $\eta=0$ and projective measurements for $\eta=1$, which was originally developed in Ref.~\cite{gruenbaum13}.
The goal is to study this interpolation and to identify relations, based on perturbation theory with respect to
the parameter $\eta$.

The choice of a weak measurement is equivalent to an evolution through a quantum phase damping
channel~\cite{nielsen10}. This channel shears the projection operators in dependence on the 
rotation angle between the ancilla 
and the quantum system. In physical terms, such a channel describes the event 
(occurring with probability $\eta$) that a photon has been scattered without loss of energy~\cite{nielsen10}.

\subsection{Unitary return amplitude}
\label{sect:unitary}

The unitary return amplitude $u_n=\langle\psi|U^n|\psi\rangle$ with $U=e^{-iH\tau}$, whose Fourier transform reads for $|z|<1$
\beq
\label{spectr_rep3b}
{\hat u}(z)=\sum_{n\ge 1}z^n u_n
=z\langle\psi|U({\bf 1}-zU)^{-1}|\psi\rangle 
,
\eeq
becomes in spectral representation for the eigenstates $\{|E_j\rangle\}$ and eigenvalues $\{E_j\}$ of $H$
with $p_j=|\langle\Psi|E_j\rangle|^2>0$
\beq
{\hat u}(z)=\sum_{j=0}^N 
\frac{p_j}{e^{iE_j\tau}/z-1}
,\ \ \
\sum_j p_j=1
.
\eeq
We note that ${\hat u}(z)$ has only poles on the unit circle $S^1$ and no zeros {\it outside} the 
unit circle, since ${\rm Re}(e^{iE_j\tau}/z-1)=(\cos\varphi_j-|z|)/|z|<0$ for $|z|>1$, where we have used $z=|z|e^{i\alpha}$ and $\varphi_j=E_j\tau-\alpha$.
${\hat u}(z)$ can be expressed as a rational function ${\hat u}=zP_{N}/D_{N+1}$ with the polynomials
\beq
\label{polynomial3}
P_{N}=\sum_{j=0}^{N}p_j\prod_{k=0, k\ne j}^{N}(e^{iE_k\tau}-z)
\ ,\ \ \
D_{N+1}=\prod_{k=0}^{N}(e^{iE_k\tau}-z)
.
\eeq
Moreover, $1+{\hat u}(z)$ obeys the relation
\beq
\label{1+u}
1+{\hat u}(z)
=-\sum_{j=0}^{N}\frac{p_j}{e^{-iE_j\tau}z-1}
=-\left[{\hat u}(1/z^*)\right]^*
,
\eeq
which has no zeros {\it inside} the unit circle, since ${\rm Re}(1+\hat{u})>0$ for $|z|<1$.
This implies that ${\rm Re}(1+x\hat{u}(z))=1-x+x(1+{\rm Re}(\hat{u}(z)))>0$ for $|z|<1$ and $0\le x\le1$.

\subsection{Monitored return and transition amplitudes}
\label{sect:monitor}

With $x:=1-\sqrt{1-\eta}$ we get from the definition $Q_\eta={\bf 1}-xP$ in Eq. (\ref{def_proj}) 
the monitored return amplitude as
\beq
\label{amplitude1}
\phi_{\eta,n}=\sqrt{\eta}\langle\psi|U(Q_\eta U)^{n-1}|\psi\rangle
\eeq
and its Fourier transform
\beq
\hat{\phi}_\eta(z)=\sum_{n\ge 1}z^n\phi_{\eta,n}
=z\sqrt{\eta}\langle\psi|U({\bf 1}-zQ_\eta U)^{-1}|\psi\rangle
\label{geom_exp0}
=z\sqrt{\eta}\langle\psi|U({\bf 1}-zU+zxPU)^{-1}|\psi\rangle
.
\eeq
Employing an expansion either in powers of $z(x-1)PU$ or in powers of $zxPU$, we obtain for
the monitored return and transition amplitudes the following results
\beq
\label{ireturn}
{\hat\phi}_\eta=\frac{\sqrt{\eta}{\hat{u}}}{1+x{\hat{u}}}
=\frac{\sqrt{\eta}{\hat{\phi}}}{1-(1-x){\hat{\phi}}}
,\ \ 
{\hat\phi}_\eta'=\frac{\sqrt{\eta}{\hat{u}'}}{1+x{\hat{u}}}
=\frac{\sqrt{\eta}{\hat{\phi}'}}{1-(1-x){\hat{\phi}}}
\eeq
with the unitary transition amplitude $\hat{u}'=z\langle\psi'|U({\bf 1}-zU)^{-1}|\psi\rangle$
and the monitored transition amplitude 
$\hat{\phi}_\eta'(z)=z\sqrt{\eta}\langle\psi'|U({\bf 1}-zQ_\eta U)^{-1}|\psi\rangle$.
These results reflect the fact that the return amplitude $\hat{\phi}_\eta(z)$ 
can be approached from the unitary side with $x=0$ and from the projective monitoring side with $x=1$ 
with their corresponding return amplitudes $\hat{u}$ and $\hat{\phi}$, respectively. 

We note that ${\hat\phi}(0)=\hat{\phi}_\eta(0)=0$, since the summation in Eq. (\ref{geom_exp0}) 
starts from $n=1$. With  Eq. (\ref{1+u}) we get for $\eta=1$
\beq
\label{phase0}
{\hat\phi}(z)=
\frac{{\hat u}(z)}{1+{\hat u}(z)}=-\frac{{\hat u}(z)}{{\hat u}^*(1/z^*)}
\eeq
and with ${\hat u}=zP_{N}/D_{N+1}$ and Eq. (\ref{polynomial3})
\beq
\label{rational_f1}
{\hat\phi}(z)=-z\prod_{k=1}^{N}\left(\frac{z-{\bar z}_k}{1-z{\bar z}_k^*}\right)
\prod_{k=0}^{N}\left(-e^{-iE_k\tau}\right)
\eeq
with the zeros $\{{\bar z}_j\}$ of $P_{N}$ and ${\bar z}_0=0$. 

Finally, we take the limit $z\to e^{i\omega}$ with real $\omega$, which leads for 
${\hat \phi}(z)$ to a 
unimodular function on the unit circle:
According to Eq. (\ref{phase0}) we can write in this limit
\beq
\label{phase_factor}
{\hat\phi}(e^{i\omega})
=-\frac{{\hat u}(e^{i\omega})}{{\hat u}^*(e^{i\omega})}
\equiv e^{i f(\omega)}
\eeq
with a real phase $f(\omega+2\pi)=f(\omega)$. This limit must be taken with care though, since the poles of 
$\hat{u}(z)$ are on $S^1$ and some poles of $\hat{\phi}(z)$ can also be located on $S^1$. 
Assuming though that the zeros of $\hat{u}(z)$ are inside the unit disk, there are no singularities on $S^1$, 
since the denominator of  $\hat{\phi}(z)$ in Eq. (\ref{rational_f1}) provides poles only outside
the unit disk~\cite{poles} . 
Thus, we can integrate $z$ over $S^1$ to obtain
\beq
\frac{1}{2\pi i}\int_{S^1}\frac{\hat{\phi}^l }{z}dz=\delta_{l,0}
,
\eeq
since $\hat{\phi}(z)$ has no poles inside the unit disk and since $\hat{\phi}(0)=0$.
Using for $\hat{\phi}_\eta$ in Eq. (\ref{ireturn}) an expansion in powers of $(1-x)e^{if(\omega)}$ , 
this implies
\beq
\sum_{n\ge1}|\phi_{\eta,n}|^2
=\frac{1}{2\pi i}\int_{S^1}  \frac{|\hat{\phi}_\eta|^2}{z}dz=\eta\sum_{l\ge 0}(1-x)^{2l}=1
.
\eeq
Thus, the monitored quantum walk returns with probability 1.
We summarize that  the phase $f(\omega)$ indicates a topological aspect because it
represents an integer winding number that is equal to the number of non-degenerate poles of
$\partial_z\log\hat{\phi}(z)$ inside the unit disk.

\subsection{Mean return time}
\label{sect:mrt}

Eq. (\ref{phase_factor}) can be used to calculate the mean number of indirect measurements 
$\langle n\rangle=\sum_{n\ge 1}n|\phi_{\eta,n}|^2$ for the return to the initial state as the 
integral
\beq
\langle n\rangle=
\frac{-i}{2\pi}\int_0^{2\pi}\hat{\phi}_\eta(e^{i\omega})^*
\partial_\omega\hat{\phi}_\eta(e^{i\omega})d\omega
=\frac{-i\eta}{2\pi}\int_0^{2\pi}\frac{e^{-if}}{1-(1-x)e^{-if}}\partial_\omega
\frac{e^{if}}{1-(1-x)e^{if}}d\omega
.
\eeq
From the expansion of both denominators in a geometric series we get
\beq
\label{MRT}
\langle n\rangle=\sum_{l_1,l_2\ge 1}l_2(1-x)^{l_1+l_2}\frac{\eta}{2\pi (1-x)^2}
\int_0^{2\pi}e^{if(l_2-l_1)}\frac{df}{d\omega}d\omega
.
\eeq
For $l_2\ne l_1$ the integral vanishes:
\[
\frac{1}{2\pi}
\int_0^{2\pi}e^{if(l_2-l_1)}\frac{df}{d\omega}d\omega
=\frac{-i}{2\pi(l_2-l_1)}
\int_0^{2\pi}\frac{de^{if(l_2-l_1)}}{d\omega}d\omega=0
\]
due to the periodicity of $e^{if (l_2-l_1)}$. For $l_2=l_1$
we use Eq. (\ref{rational_f1}) to write the Cauchy integral with the closed contour $S^1$ 
\beq
\frac{1}{2\pi}
\int_0^{2\pi}\frac{df}{d\omega}d\omega
=\frac{1}{2\pi i}\int_{S^1} \frac{d\log \hat{\phi}(z)}{dz}dz=N+1
=:n_w
,
\eeq
since the $N+1$ zeros of $P_N$ become poles inside the unit disk
after a differentation of $\log\hat{\phi}$. Degenerate poles and poles on the unit circle must be
treated separately, which reduces the winding number to $n_w<N+1$~\cite{gruenbaum13}.
Thus, we obtain from the right-hand side of Eq. (\ref{MRT})
\beq
\label{MRT1}
\langle n\rangle=
\eta n_w\sum_{l\ge 1}l(1-x)^{2(l-1)}
=\frac{n_w}{\eta}
\eeq
as the mean return time under monitoring with indirect measurements. We note that this
calculation was also presented in Ref.~\cite{heine25}.

Finally, we get a recursion relation for the return amplitude by expanding $\hat{\phi}_\eta(z)$
and $\hat{\phi}(z)$ in powers of $z$. Using the relation (\ref{ireturn}), we relate the coefficients
of $z^n$ for both expansions and get
\beq
\label{recursive_rel}
\frac{1}{\sqrt{\eta}}\phi_{\eta,n}
=\sum_{m=1}^nq_{n,m}(1-x)^{m-1}
\eeq
with
\beq
q_{n,m}=\sum_{k_1,\ldots,k_{m-1}=1}^{m-1}\phi_{k_1}\cdots \phi_{k_{m-1}}\phi_{n-k_1-\cdots -k_{m-1}}
\eeq
and $\phi_k=0$ for $k<1$.  $|\phi_n|\le 1$ with $\sum_n |\phi_n|^2=1$
and $0\le x<1$ yields a fast non-monotonic decay of higher order terms in $x$.

\subsection{Variance of the return time}
\label{sect:variance}

The variance of the return time $\langle t^2\rangle-\langle t\rangle^2 =\tau^2(\langle n^2\rangle-\langle n\rangle^2)$
can also be obtained from a Cauchy integration of the generating function $|\hat{\phi}_{\eta}(z)|^2$.
Alternatively, we can employ a spectral representation of $\phi_{\eta,n}$. To this end, we
consider the complex eigenvalues $\{\lambda_j\}$ of $Q_\eta U$, which are located on the unit disk.
Including only bright states with $|\lambda_j|<1$, we write the amplitude in Eq. (\ref{amplitude1}) as
\beq
\phi_{\eta,n}=\sqrt{\eta}\langle\psi|U(Q_\eta U)^{n-1}|\psi\rangle
=\sqrt{\eta}\sum_{j=0}^N a_j\lambda_j^{n-1}
.
\eeq
The coefficients $\{a_j\}$ are determined by the eigenvectors of $Q_\eta U$, multiplied by the initial 
state $|\psi\rangle$. This means that they do not depend on the time $\tau$ between measurements,
in contrast to the eigenvalues.
This gives us immediately the return probability
\beq
|\phi_{\eta,n}|^2
 =\eta\sum_{j,k} a_j a_k^* (\lambda_j \lambda_k^*)^{n-1}
.
\eeq
With the relations 
\beq
\partial_\gamma\sum_{n\ge1}\gamma^n
=\sum_{n\ge 1}n\gamma^{n-1}=1/(1-\gamma)^2
,\ \
\partial_\gamma^2\sum_{n\ge1}\gamma^{n+1}
=\sum_{n\ge 1}n(n+1)\gamma^{n-1}=2/(1-\gamma)^3 
\eeq
for $|\gamma|<1$, we get for the mean return time
\beq
\langle n\rangle
=\sum_{n\ge1} n |\phi_{\eta,n}|^2
=\eta\sum_{j,k=0}^{N} a_j a_k^*
\sum_{n\ge1} n(\lambda_j\lambda_k^*)^{n-1}
=\eta\sum_{j,k=0}^{N} \frac{a_j a_k^*}{(1-\lambda_j \lambda_k^*)^2}
\eeq
and for the second moment of the return time
\beq
\langle n^2\rangle
=\sum_{n\ge1} n^2|\phi_{\eta,n}|^2
=\eta\sum_{j,k=0}^{N} a_j a_k^*\sum_{n\ge1} n^2(\lambda_j\lambda_k^*)^{n-1}
=\eta\sum_{j,k=0}^{N} a_j a_k^*\frac{1+\lambda_j \lambda_k^*}{(1-\lambda_j \lambda_k^*)^3}
.
\eeq
In the unitary limit (i.e., $x=0$) the eigenvalues are $\lambda_j=e^{-iE_j\tau}$. Standard perturbation
theory in $x$ yields in first order $\lambda_j=e^{-iE_j\tau}
(1-x\langle E_j|P|E_j\rangle)+O(x^2)$. The coefficients
read in the unitary limit $a_j=\langle \Psi|E_j\rangle$, which are not affected in first order 
of $x$. Thus we obtain asymptotically for $x\sim0$
\beq
\langle n\rangle\sim
\eta\sum_{j,k=0}^{N} \frac{a_j a_k^*}
{[1-e^{-i(E_j-E_k)\tau}(1-x\langle E_j|P|E_j\rangle\langle E_k|P|E_k\rangle)]^2}
\eeq
and
\beq
\langle n^2\rangle\sim
\eta\sum_{j,k=0}^{N} a_j a_k^*\frac{1+e^{-i(E_j-E_k)\tau}}
{[1-e^{-i(E_j-E_k)\tau}(1-x\langle E_j|P|E_j\rangle\langle E_k|P|E_k\rangle)]^3}
.
\eeq
Due to
$x=1-\sqrt{1-\eta}$ we get $\eta\sim x\sim 0$. This implies that the denominators are
$\sim \eta^2$ for $\langle n\rangle$ and $\sim\eta^3$ for $\langle n^2\rangle$, such that
\beq
\label{variance2}
\langle n\rangle \sim \frac{\rho_1}{\eta}
,\ \ 
\langle n^2\rangle \sim \frac{\rho_2}{\eta^2}
\eeq
with coefficients $\rho_{1,2}$, independent of $\eta$ and $\tau$.
For $\langle n\rangle$ this agrees with the more general result of Eq. (\ref{MRT1}).

\begin{figure}[t]
    \centering
 \includegraphics[width=7cm,height=5cm]{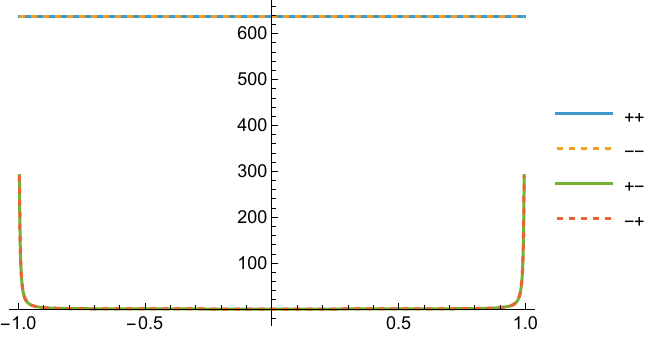}
     \caption{The variance of the return time in the 2-level system as a function of $\cos J\tau$ for $x=1/7$.
     The symbols in the legend refer to the matrix elements $|\nu_{++}|$, $|\nu_{--}|$, $|\nu_{+-}|$, and $|\nu_{+-}|$.
           }    
\label{fig:1}
\end{figure}

As an instructive example we consider the two-level system.
The two eigenvalues $\lambda_\pm$ of $Q_\eta U$ are (cf. App. \ref{app:2_level})
\begin{equation}
\lambda_\pm
=
\frac{(2-x)\cos(J\tau)
\pm
\sqrt{x^2\cos^2(J\tau)-4(1-x)\sin^2(J\tau)}}{2}
,
\end{equation}
which give a variance that is constant over a wide range of $\cos J\tau$. Since the coefficients $a_j$
are independent of $\tau$, it is sufficient to study the matrix elements
\[
\nu_{jk}=\frac{1+\lambda_j \lambda_k^*}{(1-\lambda_j \lambda_k^*)^3}
.
\]
As illustrated in Fig. \ref{fig:1}, these matrix elements are constant with respect to a changing
$\cos J\tau$, except near the singular points $\cos J\tau=\pm1$. Whether or not these variance plateaus
exist also for larger systems will be addressed in a forthcoming paper, using perturbation theory and
numerical simulations. 

\section{Discussion and conclusion}
\label{sect:discussion}

A main result of our study is that the measurement parameter $\eta$ 
appears in the renormalization of the monitored return amplitude, as well as the 
monitored transition amplitude, by the same factor $\sqrt{\eta}/[1+\sqrt{1-\eta}\hat{\phi}]$ in
Eq. (\ref{ireturn}). Since $\hat{\phi}$ is subject of a topologically preserved winding number, 
this means that the renormalization due to the measurements has a topological origin.
Another important outcome of our analysis is that the generating function $\hat{\phi}_\eta(z)$
has either a convergent expansion in powers of $x\equiv1-\sqrt{1-\eta}$ about the unitary limit $\eta=0$
or an expansion in powers of $1-x\equiv\sqrt{1-\eta}$ about the limit $\eta=1$ of projective 
measurements.  This type of expansion is applicable as a geometric series for any shearing operator 
$Q_\eta$ in Eq. (\ref{geom_exp0}).

According to Eqs. (\ref{MRT1} and (\ref{variance2}), the mean return time diverges with $\tau\rho_1/\eta$
and the fluctuations diverge asymptotically with $\tau^2\rho_2/\eta^2$, when the strength of the indirect 
measurements is reduced.
It is remarkable that the coefficients $\rho_1$, $\rho_2$ do not depend on the details of the measurement.
This reflects that the mean return time $\langle t\rangle$ as well as the variance 
$\langle t^2\rangle-\langle t\rangle^2$ depend both on the rescaled time $\tau/\eta$ as the
fundamental time scale of the measurement process, at least asymptotically for $\eta\sim0$.

For any application of our measurement protocol 
it is crucial to analyze its robustness. A source of externally caused fluctuations is the
limited accuracy of the time $\tau$ between successive measurements. 
Projective measurements at random times $\{\tau_j\}$ yield the relation \cite{kessler21,ziegler21}
\beq
\label{average_t}
\langle t\rangle=\langle\tau\rangle n_w
;
\eeq
i.e., we get the same scaling for the mean return time as for the indirect measurements in Eq. (\ref{MRT1}), 
when we replace $1/\eta$ by the
average time $\langle\tau\rangle$ between successive projective measurements. It should be
noted though that the scaling range of the random-time measurements is larger than the
corresponding range of the indirect measurements due to $0\le\langle\tau\rangle<\infty$,
in contrast to $1\le 1/\eta<\infty$.

A special case of random measurement times is a discrete distribution, which we get when there 
is a projective measurement with probability $p$ and no measurement with probability $1-p$. 
This leads to the distribution
\beq
P(n\tau)=(1-p)^{n-1}p
\eeq
of the time intervals $n\tau$ between successive measurements. For the mean time between two
successive measurements, we get $\langle\tau\rangle=\tau/p$ and with the relation in Eq. 
(\ref{average_t}) for the mean return time $\langle t\rangle=n_w/p$.
We note that this discrete random monitoring protocol can also be described by the Kraus operators
\beq
K_{0} = \sqrt{p} P U, 
\ \
K_{0,1} = \sqrt{p} Q U,
\ \
K_{0,2} = \sqrt{1-p} U .
\eeq
This implies that we replace the evolution operator $QU$ of projected measurements by the operator
$R_pU=K_{0,1}+K_{0,2}=[(\sqrt{p}+\sqrt{1-p}){\bf 1}-\sqrt{p}P]U$.
Then, instead of the expression in Eq. (\ref{geom_exp0}), we have
\beq
\hat{\phi}_{p}(z)
=\sum_{n\ge 1}z^n\phi_{p,n}
=\sqrt{p}z\langle\psi|U({\bf 1}-zR_p U)^{-1}|\psi\rangle
\eeq
with $R_p=(\sqrt{p}+\sqrt{1-p}){\bf 1}-\sqrt{p}P$. The square roots of the probabilities $p$ and $1-p$
appear, since we average the return probability $|\hat{\phi}_{p}(z)|^2$, for which always two factors appear for the 
same time~\cite{ziegler21}. This result gives the same return amplitude as for 
indirect measurements, where only $Q_\eta$ is replaced by $R_p$ now.

\begin{figure}[t]
    \centering
 \includegraphics[width=8cm,height=6cm]{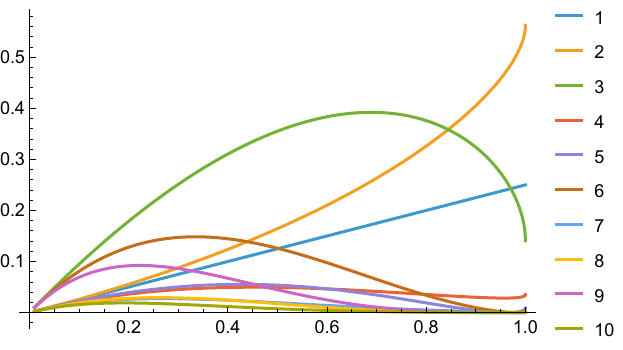}
     \caption{The return probabilities $|\phi_{\eta,n}|^2$ of a single qubit for $\cos J\tau=1/2$ 
     and $n=1,\dots,10$ as a function of $\eta$.
           }    
\label{fig:2}
\end{figure}

Relation (\ref{recursive_rel}) provides a practical tool for a systematic approximation
of the monitored return probability $|\phi_{\eta,n}|^2$ due to the fast decay of $q_{n,m}$ with $m$.
It is useful for applications to quantum circuits and quantum algorithms, where only a relatively small 
number of measurements occur. 
Although the mean return time obeys a simple scaling law with $1/\eta$, the return probabilities
$|\phi_{\eta,n}|^2$ in Eq. (\ref{recursive_rel}) do not indicate any scaling behavior. This can
already be seen for the case of a single qubit in Fig. \ref{fig:2}, where the $\eta$ dependence
varies significantly for different values of $n$ (cf. details in App. \ref{app:2_level}). In particular,
some return probabilities are increased by larger $\eta$, others are decreased.

The decay of probabilities $|\phi_k|^2$ and $|\phi_{\eta,k}|^2$ for $k\ge 1$
can be studied with the help of the eigenvalues of the matrices $QU$ and $Q_\eta U$,
respectively, with $|\lambda_j|\le1$ as discussed in Sect. \ref{sect:variance}. 
They are also the inverse poles of the generating function $z\sqrt{\eta}\langle\psi|U({\bf 1}-zQ_\eta U)^{-1}|\psi\rangle$.
We have assumed a pure state for the projection $P=|\psi\rangle\langle\psi|$. An important consequence is
that the geometric series in Eq. (\ref{geom_exp0}) appears with powers of the scalar 
$\hat{u}=\langle\psi|U({\bf 1}-zU)^{-1}|\psi\rangle$.
For a mixed state $P=\sum_j P_j|\psi_j\rangle\langle\psi_j|$ the geometric series would give
products of matrices with off-diagonal matrix elements $\langle\psi_j|U({\bf 1}-zU)^{-1}|\psi_k\rangle$.
Thus, even in this more general case we can define a return amplitude in a geometric series, which 
would be ruled by integer winding numbers. However,
there can be different winding numbers for different eigenvalues, such that the mean time may not 
be quantized. 

\appendix

\section{
Single qubit (2-level system)}
\label{app:2_level}

The monitored 2-level system is an example to test the general concept of weak measurements.
It was previously analyzed in a similar context~\cite{gebhart20}, 
we will use it here to derive explicit expressions for the monitored amplitudes of Sect. \ref{sect:monitor}.
For this purpose we consider the eigenstates of the two-level Hamiltonian $H$: 
$H|\u\rangle=J|\u\rangle$ and $H|\d\rangle=-J|\d\rangle$ and introduce the transformed basis
states $\{|\Psi\rangle,|\Psi'\rangle\}$ with
\beq
|\Psi\rangle=\frac{1}{\sqrt{2}}(|\u\rangle+|\d\rangle)
, \ \ 
|\Psi'\rangle=\frac{1}{\sqrt{2}}(|\u\rangle-|\d\rangle)
.
\eeq
Thus, we can write
\beq
\langle\Psi|e^{-iH\tau}|\Psi\rangle=\langle\Psi'|e^{-iH\tau}|\Psi'\rangle=\cos J\tau
,
\eeq
\beq
\langle\Psi'|e^{-iH\tau}|\Psi\rangle=
\langle\Psi|e^{-iH\tau}|\Psi'\rangle=-i\sin J\tau
.
\eeq
The monitored return amplitude reads with $U=e^{-iH\tau}$
\beq
\phi_k=\langle\Psi|U(QU)^{k-1}|\Psi\rangle
=\langle\Psi|U|\Psi'\rangle\langle\Psi'|U|\Psi'\rangle\cdots\langle\Psi'|U|\Psi\rangle
=\begin{cases}
\cos J\tau & k=1 \\
-\sin^2(J\tau)\cos^{k-2}(J\tau) & k\ge 2
\end{cases}
\eeq
and the monitored transition amplitude
\beq
\phi_k'=\langle\Psi'|U(QU)^{k-1}|\Psi\rangle
=\langle\Psi'|U|\Psi'\rangle\langle\Psi'|U|\Psi'\rangle\cdots
\langle\Psi'|U|\Psi\rangle
=-i\cos^{k-1}(J\tau)\sin J\tau
\eeq
and the corresponding probabilities for $J\tau=0\ ({\rm mod}\ \pi)$ are
\beq
|\phi_k|^2=\delta_{k1}
,\ \ 
|\phi_k'|^2=0
;
\eeq
i.e., $|\Psi'\rangle$ is a dark state in this case.
This result yields immediately with $|z|<1$
\beq
\label{FT01}
\hat{\phi}(z)=\sum_{k\ge 1}z^k\phi_k=z\frac{\cos J\tau-z}{1-z\cos J\tau}
,
\eeq
which implies $|\hat{\phi}(e^{i\omega})|=1$ and has one pole at $1/\cos J\tau$. 
Here we must exclude the points $J\tau=0\ ({\rm mod}\ \pi)$,
where $\hat{\phi}(z)$ has poles on $S^1$. 
Moreover, for the transition amplitude we get
\beq
\label{FT02}
\hat{\phi}'(z)=\sum_{k\ge 1}z^k\phi_k'=\frac{-iz\sin J\tau}{1-z\cos J\tau}
.
\eeq
These results can be used to write for the monitored amplitudes
%
%
\beq
\hat{\phi}_\eta(z)=\frac{\sqrt{\eta}z(\cos J\tau -z)}{1+\sqrt{1-\eta}z^2-z(1+\sqrt{1-\eta})\cos J\tau}
,
\eeq
\beq
\hat{\phi}'_\eta(z)=\frac{-i\sqrt{\eta}z\sin J\tau}{1+\sqrt{1-\eta}z^2-z(1+\sqrt{1-\eta})\cos J\tau}
.
\eeq
We note that both functions vanish at $z=0$, reflecting that the summations in Eqs. (\ref{FT01})
and (\ref{FT02}) start at $k=1$.
There is one pole at $|z|=1$ for $J\tau=0\ ({\rm mod}\ \pi)$, while for other values of $J\tau$ the two poles 
are outside the unit disk. Moreover, the zeros are inside the unit disk except for $J\tau=0\ ({\rm mod}\ \pi)$.

\vskip0.5cm

\noindent
{\bf Acknowledgment:}
We are grateful to Eli Barkai for interesting discussions in the early stage of this work.


\begin{thebibliography}{99}
\bibitem{poles}
If there are factors in Eq.~(\ref{rational_f1}) with $|\bar{z}_k|=1$, they can be removed from the product.
This means that we project the system onto one that is governed by less than $N+1$ poles.
\end{thebibliography}
\end{document}